\documentclass[graybox]{svmult}

\usepackage{mathptmx}       
\usepackage{helvet}         
\usepackage{courier}        
\usepackage{type1cm}        
\usepackage{makeidx}         
\usepackage{graphicx}        
\usepackage{multicol}        
\usepackage[bottom]{footmisc}
\usepackage{amssymb}
\usepackage{amsmath}
\usepackage{epsf}

\makeindex             

\begin{document}

\title*{Magic Numbers in the Discrete Tomography of Cyclotomic Model Sets}
\titlerunning{Magic Numbers in Discrete Tomography}
\author{Christian Huck}
\institute{Christian Huck \at Fakult\"at f\"ur Mathematik,
  Universit\"at Bielefeld, Postfach 100131, 33501 Bielefeld, Germany, \email{huck@math.uni-bielefeld.de}}
%
%
\maketitle

\abstract*{We report recent progress in the problem of distinguishing 
  convex subsets of cyclotomic model sets $\varLambda$ by
  (discrete parallel) X-rays in prescribed
  $\varLambda$-directions. It turns out that for any of these model sets
  $\varLambda$ there exists a `magic number' $m_{\varLambda}$ such
  that any two convex subsets of $\varLambda$  can be distinguished by
  their X-rays in any set of $m_{\varLambda}$ prescribed
  $\varLambda$-directions. In particular, for pentagonal, octagonal,
  decagonal and dodecagonal model sets, the least possible numbers are in that very order 11, 9, 11 and 13.}

\abstract{We report recent progress in the problem of distinguishing 
  convex subsets of cyclotomic model sets $\varLambda$ by
  (discrete parallel) X-rays in prescribed
  $\varLambda$-directions.  It turns out that for any of these model sets
  $\varLambda$ there exists a `magic number' $m_{\varLambda}$ such
  that any two convex subsets of $\varLambda$  can be distinguished by
  their X-rays in any set of $m_{\varLambda}$ prescribed
  $\varLambda$-directions. In particular, for pentagonal, octagonal,
  decagonal and dodecagonal model sets, the least possible numbers are in that very order 11, 9, 11 and 13.}

\section{Introduction}
\label{sec:1}
{\em Discrete tomography} is concerned with the 
inverse problem of retrieving information about some {\em finite}
set in Euclidean space from 
(generally noisy) information about its slices. One important problem
is the {\em unique reconstruction} of a finite point set in Euclidean $3$-space 
from its {\em (discrete parallel) X-rays} in a small number of directions, where the
  {\em X-ray} of the finite set in a certain direction is the {\em line sum
  function} giving the
number of points in the set on each line parallel to this direction. 

The interest in the discrete tomography of planar Delone sets $\varLambda$
with long-range order is motivated by the requirement in materials
science for the unique reconstruction of solid state materials like {\em quasicrystals}
slice by slice from their images under quantitative {\em high
  resolution transmission electron microscopy} (HRTEM). In fact,
in~\cite{ks}, \cite{sk} a technique
is described, which can, for certain 
crystals, effectively measure the number of atoms lying on densely occupied 
columns. 
Clearly, the aforementioned density condition forces us to consider
only $\varLambda$-directions, i.e.\ directions parallel to lines through
two different points of $\varLambda$. 

In the 
 {\em quasicrystallographic} setting,  the
 positions to be determined form a finite subset of a nonperiodic Delone
   set with long-range order 
 (more precisely, a
{\em model set}~\cite{BM}, \cite{Moody}). Model sets
 possess a dimensional hierarchy, i.e.\ they allow a slicing into
 planar model sets. In fact, many of
 the model sets that describe real quasicrystallographic structures
 allow a slicing such that each slice is an {\em $n$-cyclotomic model set}, the
 latter being (planar) Delone sets contained in the additive subgroup
 of the Euclidean plane 
 generated by the $n$th roots of unity; cf.~\cite{H},
 \cite{H2}, \cite{H5} and \cite{St}
 for details. It therefore suffices to study the discrete tomography
 of these cyclotomic
 model sets. In practice, the cases $n=5,8,12$ are of particular
 interest. In the present text, we shall mainly focus
 on the larger class of {\em cyclotomic Delone sets}.

Since different finite subsets of an $n$-cyclotomic model set $\varLambda$ may have the same X-rays in several
directions, 
one is naturally interested in conditions
to be imposed
on the set of directions together with restrictions on
the possible finite subsets of
$\varLambda$ such that the latter phenomenon
cannot occur. Here, we consider the {\em convex subsets}
of $\varLambda$ and summarise recent results in the problem of
distinguishing convex subsets of  $\varLambda$ by X-rays in prescribed
$\varLambda$-directions. It turns that there are four prescribed $\varLambda$-directions such that any two convex subsets of
$\varLambda$ can be distinguished by the corresponding X-rays,
whereas less than four $\varLambda$-directions never suffice for this
purpose. Much more novel is
the result obtained in collaboration with M.\ Spie{\ss} that there is a finite number
$m_{\varLambda}$ such that any two convex subsets of $\varLambda$
can be distinguished by their X-rays in {\em any} set of
$m_{\varLambda}$ prescribed
$\varLambda$-directions. Moreover, the
least possible numbers $m_{\varLambda}$ in the
case of the practically most relevant examples of $n$-cyclotomic model
sets $\varLambda$ with $n=5$, $8$ and $12$ only depend on $n$ and are in that
 very order $11$, $9$ and $13$. This extends a well-known result of R.\  J.\ Gardner
and P.\ Gritzmann~\cite{GG} on the corresponding
problem for the crystallographic cases $n=3,4$ of the triangular
resp.\ square 
lattice $\varLambda$ (with least possible number $m_{\varLambda}=7$ in
both cases) to
cases that are relevant in quasicrystallography. 

The intention of this text is to provide an easy to read guide to the
results of~\cite{HS} with a view towards practical applications. Detailed proofs, related results and an
extensive list of references can be found there. For the algorithmic 
reconstruction problem in the quasicrystallographic setting, the
reader is referred to~\cite{BG2},~\cite{H2}.

\section{Cyclotomic Delone sets}
\label{sec:2}

Throughout the text, the Euclidean plane is 
identified with the complex numbers. For $z\in\mathbb C$, $\bar{z}$ denotes the 
complex conjugate of $z$. Further, we denote by $K_{\varLambda}$
the smallest subfield of $\mathbb C$ that contains the rational
numbers as well as the union 
of $\varLambda-\varLambda$ 
and its image $\overline{\varLambda-\varLambda}$ under complex
conjugation. Recall that $\varLambda$ is called a {\em Delone set} if it is both uniformly
  discrete and relatively dense. For $n\in \mathbb{N}$, we always let $\zeta_n = e^{2\pi
    i/n}$, a primitive $n$th root of unity in $\mathbb C$. Then, the smallest
  subfield of $\mathbb C$ that contains the rational
numbers as well as $\zeta_n$  is the
  $n$th cyclotomic field denoted by $\mathbb Q(\zeta_n)$. The latter
  ist just the $\mathbb Q$-span of the $n$th roots of unity and
  thus contains the $\mathbb Z$-span $\mathbb Z[\zeta_n]$ of the $n$th roots of unity.  Recall
  that a {\em homothety}\/ of the complex plane  is given by $z \mapsto \lambda z + t$, where
$\lambda \in \mathbb R$ is positive and $t\in \mathbb C$. For the 
purpose of this text, the following rather abstract definition provides a convenient framework.   

\begin{definition}\label{algdeldef}
Let $n\geq 3$. A Delone set $\varLambda\subset\mathbb C$ is called an {\em $n$-cyclotomic
  Delone set} if it satisfies the following properties:
\begin{eqnarray*}
\mbox{($n$-Cyc)}&&K_{\varLambda}\mbox{ is contained in } \mathbb Q(\zeta_n)\,.\\
\mbox{(Hom)}&&\mbox{For any finite subset $F$ of $K_{\varLambda}$, there is a
homothety $h$ of}\\&&\mbox{the complex plane that maps the
elements of $F$ to $\varLambda$\,. 
}
\end{eqnarray*}
Further,
$\varLambda$ is called a {\em cyclotomic Delone set} if it is an
$n$-cyclotomic Delone set for a suitable $n\geq 3$.
\end{definition} 

Standard examples of $n$-cyclotomic Delone sets are
the {\em $n$-cyclotomic
  model sets},  which were also called {\em cyclotomic model sets with underlying
$\mathbb Z$-module $\mathbb Z[\zeta_n]$}
in~\cite[Section~4.5]{H5} and are defined via the canonical
cut and project scheme that is given by the Minkowski
representation of  the $\mathbb Z$-module $\mathbb Z[\zeta_n]$; see~\cite{H5,HS} for details. These sets are certain Delone subsets of
the $\mathbb Z$-module $\mathbb Z[\zeta_n]$ and   
range from periodic examples like the fourfold square lattice ($n=4$)
or the sixfold triangular lattice ($n=3$) to nonperiodic
examples like the vertex set of the tenfold T\"ubingen triangle
tiling ($n=5$), the eightfold Ammann-Beenker tiling ($n=8$) or the twelvefold
shield tiling ($n=12$);
see~\cite[Figure~1]{H4}, ~\cite[Figure~2]{H5} and Figure~\ref{fig:tilingupolygon} below for
illustrations. Note that  the vertex sets of the famous Penrose tilings of the
plane fail to be $5$-cyclotomic model sets but can still be seen to be $5$-cyclotomic Delone sets; see~\cite{bh}
and references therein.

\section{Determination of convex subsets by X-rays}
\label{sec:3}

Let $(t_1,t_2,t_3,t_4)$ be an ordered tuple of four distinct
elements of $\mathbb{R}\cup\{\infty\}$. Then, its {\em cross ratio}\/
$\langle t_1,t_2,t_3,t_4\rangle$ is the nonzero real number defined by
$$
\langle t_1,t_2,t_3,t_4\rangle = \frac{(t_3 - t_1)(t_4 - t_2)}{(t_3 - t_2)(t_4 - t_1)}\,,
$$
with the usual conventions if one of the $t_i$ equals
$\infty$. 

The
unit circle in $\mathbb C$ is
denoted by $\mathbb{S}^{1}$ and its elements are also called
{\em directions}. For a nonzero complex number $z$, we denote
by $\operatorname{sl}(z)$ the slope of $z$, i.e.\ $\operatorname{sl}(z)=-i(z-\bar{z})/(z+\bar{z})\in
\mathbb{R}\cup\{\infty\}$. Let $\varLambda$ be a subset of
$\mathbb C$. A direction
$u\in\mathbb{S}^{1}$ is called a {\em $\varLambda$-direction} if it is
parallel to a nonzero element of the {\em difference set}\/ 
$\varLambda-\varLambda=\{v-w\,|\, v,w\in\varLambda\}$ of
$\varLambda$. By construction, the
cross ratio of slopes of four pairwise nonparallel
$\varLambda$-directions is an element of the field
$K_{\varLambda}\cap\mathbb R$. In case of $n$-cyclotomic Delone
sets $\varLambda$, these cross ratios are thus elements of the field
$\mathbb Q(\zeta_n)\cap\mathbb R$.

\begin{definition}\label{xray..}
\begin{itemize}
\item
Let $F$ be a finite subset of $\mathbb C$, let $u\in
\mathbb{S}^{1}$ be a direction, and let $\mathcal{L}_{u}$ be the set
of lines in the complex plane in direction $u$. Then the {\em
  (discrete parallel)}\/ {\em X-ray} of $F$ {\em in direction} $u$ is
the function $X_{u}F: \mathcal{L}_{u} \rightarrow
\mathbb{N}_{0}=\mathbb{N} \cup\{0\}$, defined by $$X_{u}F(\ell) =
\operatorname{card}(F \cap \ell\,)\,.$$
\item
Let $\mathcal{F}$ be a collection of finite subsets of
$\mathbb C$ and let $U\subset\mathbb{S}^{1}$ be a finite set
of directions. We say that the elements
of $\mathcal{F}$ are {\em determined} by the X-rays in the directions
of $U$ if different elements of $\mathcal{F}$ cannot have the same
X-rays in the directions of $U$.
\end{itemize}
\end{definition}

\begin{figure}
\centerline{\epsfysize=0.52\textwidth\epsfbox{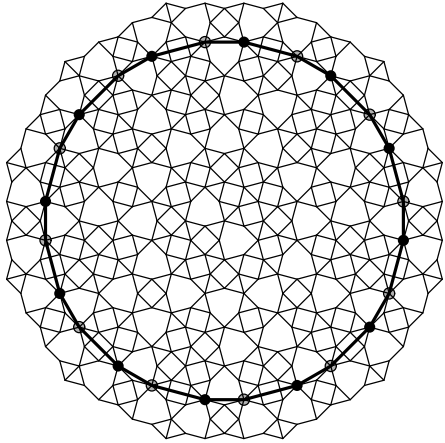}}
\caption{The boundary of a $U$-polygon in
  the vertex set $\varLambda$ of the twelvefold 
  shield tiling, where $U$ is the set of
  twelve pairwise nonparallel $\varLambda$-directions
 given by the edges and diagonals of the central regular
 dodecagon. The vertices of $\varLambda$ in the interior of
 the  $U$-polygon together with the vertices indicated by the black and grey dots, respectively,
give two different convex subsets of $\varLambda$ with the same X-rays in the
directions of $U$.}
\label{fig:tilingupolygon}
\end{figure}

One can easily see that no finite set of pairwise nonparallel
$\varLambda$-directions suffices in order to determine the whole class
of finite subsets of $\varLambda$  by the corresponding X-rays~\cite{H5}. It is
therefore necessary to impose some restriction on the finite subsets of
$\varLambda$ to
be determined. It has proven most fruitful to focus on the {\em convex
  subsets}\/ of cyclotomic Delone
sets. The latter are bounded (and thus finite) subsets $C$ of $\varLambda$ satisfying
the equation $C =
\operatorname{conv}(C)\cap \varLambda$, where $\operatorname{conv}(C)$
denotes the convex hull of $C$. One has the following fundamental 
result which shows that one has to choose the set $U$ of
$\varLambda$-directions in such a way that certain convex polygons
cannot exist; cf.~\cite[Proposition~4.6 and Lemma~4.5]{H5}. Here, for  a finite set $U\subset \mathbb{S}^{1}$ of
directions, a nondegenerate convex polygon $P\subset\mathbb C$ is
called a {\em $U$-polygon} if it has the property that whenever $v$ is
a vertex of $P$ and $u\in U$, the line in the complex plane in direction $u$
which passes through $v$ also meets another vertex $v'$ of $P$. $P$ is
called a 
{\em $U$-polygon in $\varLambda$}, if its vertices lie in
$\varLambda$. Note
that the proof of direction
(ii)$\Rightarrow$(i) needs property
(Hom) and see Figure~\ref{fig:tilingupolygon} for an
illustration of the other direction
(i)$\Rightarrow$(ii).

\begin{theorem}\label{characungen}
Let $\varLambda$ be a cyclotomic Delone set and let $U\subset\mathbb{S}^{1}$ be a set of two or more pairwise nonparallel $\varLambda$-directions. The following statements are equivalent:
\begin{enumerate}
\item[(i)]
The convex subsets of $\varLambda$ are determined by the X-rays in the directions of $U$.
\item[(ii)]
There is no $U$-polygon in $\varLambda$.
\end{enumerate}
In addition, if $\operatorname{card}(U)<4$, then there is a $U$-polygon in $\varLambda$. 
\end{theorem}

Employing Darboux's theorem on second 
midpoint polygons~\cite{G} 
together with a blend of sophisticated methods from the theory of
cyclotomic fields and previous results obtained by Gardner and Gritzmann~\cite{GG}, one obtains the
following deep result on $U$-polygons; cf.~\cite[Theorem 5.7]{HS}. Note that the proof heavily relies on
property ($n$-Cyc).

\begin{theorem}\label{finitesetncr0gen}
Let $n\geq 3$ and let $\varLambda$ be an $n$-cyclotomic Delone set. Further, let $U\subset\mathbb{S}^1$ be a set of four or more pairwise
nonparallel $\varLambda$-directions and suppose the existence of a $U$-polygon. Then the
cross ratio of slopes of any four directions of $U$, arranged in order
of increasing angle with the positive real axis, is an element of the
finite set $\mathcal{C}_{\operatorname{lcm}(2n,12)}(\mathbb Q(\zeta_n)\cap\mathbb R)
$ of numbers in the field $\mathbb Q(\zeta_n)\cap\mathbb R$ that can be written
in the form
$$
\frac{\Big(1-\zeta_{\operatorname{lcm}(2n,12)}^{k_1}\Big)\Big(1-\zeta_{\operatorname{lcm}(2n,12)}^{k_2}\Big)}{\Big(1-\zeta_{\operatorname{lcm}(2n,12)}^{k_3}\Big)\Big(1-\zeta_{\operatorname{lcm}(2n,12)}^{k_4}\Big)}\,,
$$
where $(k_1,k_2,k_3,k_4)$ is an element of the set
$$
\big\{(k_1,k_2,k_3,k_4)\in \mathbb{N}^4 \,\big|\, k_3<k_1\leq
k_2<k_4\leq \operatorname{lcm}(2n,12)-1 \mbox{ and } k_1+k_2=k_3+k_4\big\}
\,.
$$
Moreover, $\operatorname{card}(U)$ is bounded above by a finite number
$b_n\in\mathbb N$ that only depends on $n$. In particular, one can choose
$b_3=b_4=6$, $b_5=10$, $b_8=8$ and $b_{12}=12$. 
\end{theorem}

Theorems~\ref{characungen} and ~\ref{finitesetncr0gen} now immediately imply our main result on the
determination of convex subsets of cyclotomic Delone sets; cf.~\cite[Theorem 5.11]{HS}.

\begin{theorem}\label{dtmain2}
Let $n\geq 3$ and let $\varLambda$ be an $n$-cyclotomic Delone set.
\begin{enumerate}
\item[(a)]
There are sets of four pairwise nonparallel $\varLambda$-directions
such that the convex subsets of $\varLambda$ are determined by the corresponding
X-rays. In addition, less than four pairwise nonparallel $\varLambda$-directions never
suffice for this purpose.
\item[(b)]
There is a finite number $m_n\in\mathbb N$ that only depends on $n$ such that the convex subsets of
$\varLambda$ are determined by the X-rays in any set of $m_n$
pairwise nonparallel $\varLambda$-directions. In particular, one can
choose $m_3=m_4=7$, $m_5=11$, $m_8=9$ and $m_{12}=13$. 
\end{enumerate}
\end{theorem}

\begin{svgraybox}
By Theorems~\ref{characungen} and ~\ref{finitesetncr0gen} above, it
suffices for Part~(a) to take any set of four pairwise
nonparallel $\varLambda$-directions such that the cross ratio of their
slopes, arranged in order
of increasing angle with the positive real axis, is {\em not} an element of
the finite set $\mathcal{C}_{\operatorname{lcm}(2n,12)}(\mathbb
Q(\zeta_n)\cap\mathbb R)$; cf.~\cite[Corollary 4.10]{HS} for concrete
results in the practically most important cases $n=5,8,12$
of
quasicrystallography.
\end{svgraybox}

\section{Concluding remarks}
\label{sec:4}

Our above analysis heavily relies on the assumption of ideal data and
is therefore only a very first step towards a satisfactory
tool for materials science. Further, it would certainly be interesting
to abandon the slice by slice approach and work, for a Delone set
$\varLambda$ in Euclidean $3$-space, with
$\varLambda$-directions {\em in general position} instead; compare the
approach to 3D reconstruction of atomic arrangements presented in~\cite{Saitoh}. 
In that case, it might well be that {\em seven} is a {\em universal}
magic number for the determination of convex subsets by X-rays;
cf.~\cite[Problem 2.1]{G}.

\begin{acknowledgement}
This work was supported by the German Research Council (Deutsche
For\-schungsgemeinschaft), within the CRC 701.
\end{acknowledgement}

\end{document}